\documentstyle[prb,aps]{revtex}
\input epsfig.sty
\begin{document}

\advance\textheight by 0.2in
\twocolumn[\hsize\textwidth\columnwidth\hsize\csname@twocolumnfalse%
\endcsname

\draft
\title{ {\tt to appear in Phys. Rev. B } \\
Diluted Josephson-junction arrays in a magnetic field:
phase coherence and vortex glass thresholds.}
\author{Mourad Benakli $^a$, Enzo Granato $^b$, Subodh R. Shenoy $^a$, and
Marc Gabay$^c$}
\address{$^a$ Condensed Matter Physics Group, \\
International Centre for Theoretical Physics, \\
34100 Trieste, Italy \\
$^b$ Laborat\'orio Associado de Sensores e Materiais,\\
Instituto Nacional de Pesquisas Espaciais, \\
12225 - S\~ao Jos\'e dos Campos, SP, Brasil \\
$^c$Laboratoire de Physique des Solides
Laboratoire associ\'e au CNRS, Universit\'e de Paris-Sud,\\
B\^atiment 510, 91405 Orsay Cedex, France\\ 
}
\maketitle

\begin{abstract}
The effects of random dilution of junctions on a two-dimensional
Josephson-junction array in a magnetic field are considered. For
rational values of the average flux quantum per plaquette $f$, the
superconducting transition temperature vanishes, for increasing dilution,
at a critical value $x_S(f)$, while the vortex ordering remains stable up
to $x_{VL}>x_S$, much below the value $x_p$ corresponding to the geometric
percolation threshold. For $ x_{VL}<x<x_p$ , the array behaves as a
zero-temperature vortex-glass.  Numerical results for $f=1/2$ from defect
energy calculations are presented which are consistent with this scenario.
\end{abstract}


]

Vortex glass states in disordered three-dimensional superconductors
have been the focus of much recent interest
\cite{physica,fisher,john,giamarchi,dekker}.  In the absence of
screening, they are believed to have a true superconducting phase, with
vanishing linear resistivity,  at finite temperatures. By contrast, in
two dimensions, vortex glass models \cite{fisher} and experiments on
superconducting films \cite{dekker} show that vortex glass order is
destroyed at any finite temperature with a nonzero but exponentially
small resistivity. This zero-temperature vortex glass can be
characterized by a thermal correlation length exponent $\nu_T$ which
determines, for example, the current density scale, $J_{nl} \sim
T^{1+\nu_T}$, where nonlinear behavior shows up in  the current-voltage
characteristics \cite{fisher,dekker}. Recent estimates give $\nu_T \sim
2$ for various vortex glass models \cite{fisher}.

Randomly diluted Josephson junction arrays (JJA), have been used to
model disordered superconductors \cite{john,dilu,gd}. In zero field,
the superconducting transition temperature vanishes at the percolation
threshold \cite{domb} $x_p$, where $x$ is the concentration of diluted
junctions.  For $x > x_p$ there are only uncoupled finite clusters
and long-range phase coherence is destroyed.  At $x_p$, the infinite
percolating cluster shows up in the scaling behavior of the dynamic
conductivity \cite{dilu} and  nonlinear resistivity \cite{gd}. In
the presence of an external field, a diluted JJA is an experimentally
controllable model to investigate phase coherence and vortex glasses in
two dimensions. For rational values of the flux quantum per unit cell
$f$, an ordered ($x=0$) JJA has a ground state consisting of a periodic
pinned vortex lattice, with additional discrete symmetries resulting from
commensurability effects \cite{physica,teitel}. The melting of this vortex
lattice at a temperature $T_{VL}$, driven by domain-wall excitations,
competes with the superconducting transition at $T_S$ driven by the
Kosterlitz-Thouless vortex unbinding. For $f=1/2$, these transitions
either coincide or have very close transition temperatures \cite{physica},
$T_{VL} \gtrsim T_S$. Similar behavior is expected for other low
rational values of $f$. In presence of random-dilution disorder, two
natural questions arise: (i) are there two dilution thresholds, $x_{S}$
and $x_{VL}$, for phase coherence and vortex lattice order respectively?
Does a vortex-glass phase occur over a significant range $x > x_{VL}$ ?

In this work, we argue that for rational values of  $f$, the
superconducting transition temperature of the array vanishes, for
increasing dilution, at a critical value $x_S(f)$. The vortex-lattice
ordering remains stable up to $x_{VL} (f)>$ $x_S(f)$ but both values are
much below the value $x_p$ corresponding to the geometric percolation
threshold. For $x_{VL} < x < x_p$ there is a zero-temperature
vortex glass. These features are verified numerically for $f=1/2$,
using a bond-diluted frustrated XY model on a triangular lattice, and
extensive zero temperature calculations.  Domain-wall energy
calculations  gives an estimate of a wide range, $x_{VL} < x < x_p$ ,
for a zero-temperature vortex glass below the geometrical percolation
threshold $x_p=0.652$. We find $x_S$ $=$ $0.14(1)$ and   $x_{VL}$
$=$ \thinspace $0.17(1)$ consistent with the proposed scenario. In
the vortex-glass phase, $\nu_T \sim 1.9 $ , as estimated from the size
dependence of  defect energies excitations. Interestingly enough, this
estimate is very close to the value obtained for the gauge-glass model
\cite{fisher} which may suggest a common universality class.

We consider a two-dimensional Josephson-junction array in a magnetic field 
$B $ described by the Hamiltonian of a frustrated XY model 
\begin{equation}
H=-\sum_{<ij>}J_{ij}\cos (\theta _i-\theta _j-A_{ij})
\end{equation}
where $\theta _i$ is the phase of the condensate wave function in a
grain at site $i$ and $J_{ij}$ is the Josephson coupling. The summation
is taken over all nearest neighbors of a regular reference lattice. The
dimensionless line integral of the vector potential $A_{ij}$ about each
elementary reference-lattice plaquette of area $S$ is $\sum_pA_{ij}=2\pi
f$, where the frustration parameter $ f=BS/\Phi _o$ measures the number
of flux quanta $\Phi _o$ per plaquette. A bond-dilution concentration
$x$ corresponds to $J_{ij}$ being  zero or $J$, with probabilities $x$
and $1-x$, respectively.  Since any closed loops of nonzero bonds $J$
have an area which is an integer multiple of the elementary area $S$,
the properties of this model are periodic in $f$ with period $1$, and
it is therefore sufficient to consider $0<f<1$.

\begin{figure}[tbp]
\centering\epsfig{file=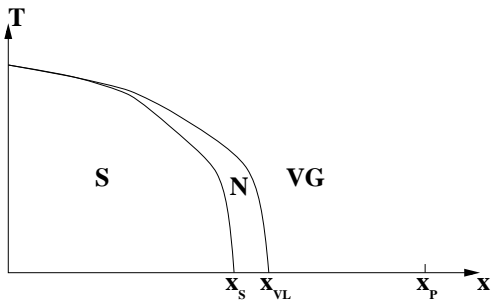,bbllx=3cm,bblly=3cm,bburx=18cm,
bbury=25cm,angle=-90,width=8cm}
\caption{Schematic phase diagram of a diluted JJA as
a function of temperature $T$ and dilution $x$, for an average rational
frustration $f$, showing a superconducting phase (S)
a normal ordered vortex-lattice phase (N) and a vortex-glass phase
(VG). A single transition is assumed at $x=0$. If the transitions  are 
already separated at $x=0$, dilution would further increase the separation.
At $T=0$, a vanishing linear resistance is expected  for $x < x_p$ 
due to vortex pinning.}  
\end{figure}

For $f=0$, the Hamiltonian reduces to the standard diluted XY model, which
is known to be superconducting for \cite{domb} $x < x_p$. When $f\ne 0$,
there must also be a threshold $x_{VL}$ for vortex-lattice disordering
\cite{giamarchi} below the percolation threshold, $x_{VL} < x_p$.  In the
undiluted case $x=0$, the ground state for rational $f=p/q$ ($q\ge 2$)
consists of a pinned vortex lattice \cite{teitel} with a $q\times q$
unit cell. For small dilution $x<<x_{VL}$, the long-range order of
the vortex lattice persists, provided an infinite cluster of these
cells exist. Since $x_{VL}(f)$ corresponds roughly to the percolation
threshold for cells of size $q\times q$, the percolation threshold for
$q\times q$ cell dilutation is reached much below the unit bond-dilution
threshold. Alternatively, long-range order of the vortex lattice requires
connectivity over at least $q\,$ bonds, as in bootstrap percolation \cite
{adler}, which is known to lead to a percolation threshold below the unit
bond percolation. Since vortex lattice disordering leads to suppression
of  phase coherence \cite{physica}, $x_{VL}$  is an upper bound for
the superconducting threshold $x_S$. This implies that the transition
temperature should vanish at an $x_S \le x_p$, and that the thresholds
are as illustrated in Fig. 1. At least for low-order rational values of
$f$, we would expect $x_S(f^{\prime })<x_S(f)$ if $f^{\prime }<f$ since
$f^{\prime}$ requires a higher connectivity. For $x_{VL}(f)<x<x_p$,
there is no long-range order, and this phase should correspond to a
two-dimensional vortex glass,  where a true phase transition is known to
occur only at $T=0$ \cite{fisher,freez}. An intervening glass phase near
percolation threshold is also expected from mean field theory \cite{john}.
This phase can be characterized \cite{fisher} by a critical exponent
$\theta $ that determines how low-energy excitations $\Delta E(L)$
from the ground state behave at long length scales $L$. For a $T=0$
vortex glass $\Delta E\sim L^\theta $, with $\theta <0$, and thermal
excitations of scale $\xi \sim T^{-\nu_T}$ destroy the glass order at
any finite temperature, leading to an identification of the thermal
correlation length exponent as $\nu _T=1/|\theta |$. Our numerical
results for $f=1/2$, described below, are consistent with this behavior,
and provide an estimation of $\nu _T$. In absence of thermal fluctuations,
at $T=0$, vortices are pinned by disorder and a nonlinear response to an
applied current is expected leading to a vanishing linear resistance and 
nonzero critical current for $ x < x_p $. 

\begin{figure}[tbp]
\centering\epsfig{file=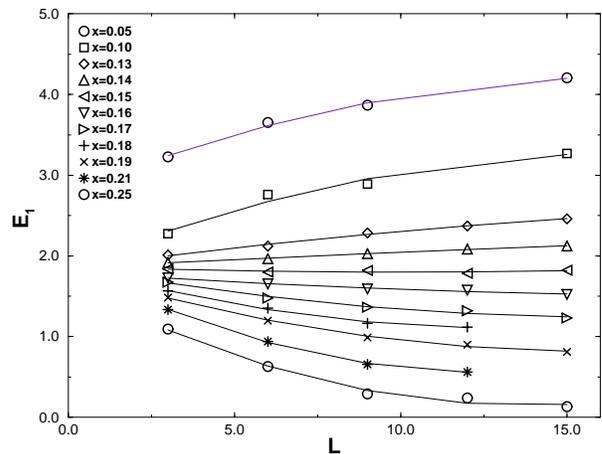,bbllx=3cm,bblly=3cm,bburx=20cm,
bbury=25cm,angle=-90,width=8cm}
\caption{Finite size behavior of defect energy $[E_1]$ probing the 
superfluid density for increasing dilution $x$ and various system sizes $L$.
The change in the $L$ dependence determines the threshold $x_S$.}
\end{figure}

\begin{figure}[tbp]
\centering\epsfig{file=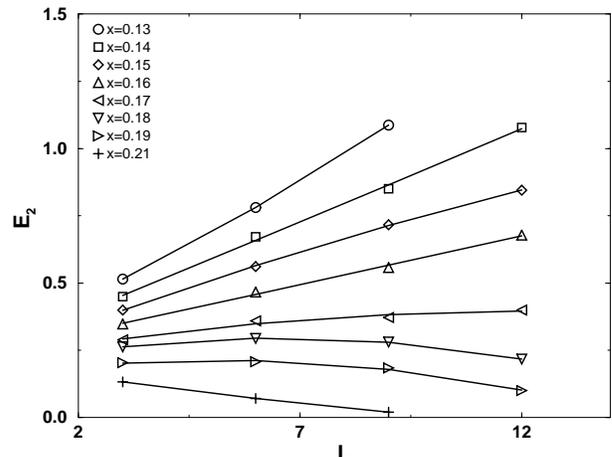,bbllx=3cm,bblly=3cm,bburx=20cm,
bbury=25cm,angle=-90,width=8cm}
\caption{Finite size behavior of defect energy $[E_2]$ probing the 
vortex-lattice lattice stability for increasing dilution $x$ and various system sizes $L$. The change in $L$ dependence determines the threshold $x_{VL}$.}
\end{figure}

We have carried out a detailed numerical study for $f=1/2$ at zero
temperature, using a bond diluted frustrated JJA on a triangular lattice,
where the critical dilution threshold for bond percolation \cite{domb}
is $x_p $ $=$ $0.652$. For this value of $f$, vortex-lattice ordering can
be conveniently described in terms of a $Z_2$ chirality order parameter
$\chi =\sum_{<ij>}(\theta _i-\theta _j-A_{ij})/(2\pi )$, where summation
is taken about an elementary plaquette of the actual lattice, and the
gauge-invariant phase difference is restricted to the interval $[-\pi
,+\pi ] $. In the undiluted case, the ground state consists of a pinned
vortex-lattice corresponding to an antiferromagnetic arrangement of
$\chi =\pm 1/2$. To study the stability of the ordered phases , we use
a defect energy renormalization analysis \cite {bray} at $T=0$. A defect
is created in a system of size $L$ $\times $ $L\,$ by imposing a change
in the boundary conditions in one direction. The change  $\Delta E(L)$
in the ground state energy for small systems is calculated for a large
number of samples by directly searching for the minimum energy. We used
an improved algorithm based on Ref. \onlinecite{min}. Typically, $3000$
configurations of disorder have been used for each system size. To study
both phase coherence and vortex-lattice order, we consider two types
of defects: (i) From the energy difference between periodic $E_p$ and
antiperiodic $E_a$ boundary conditions in the phases $\theta _i$ we obtain
$\Delta E_{1} = E_a$ $-$ $E_p$, which is a measure of phase coherence,
and is related to the renormalized stiffness constant $J(L)=\rho $
$\Delta E_1/2\pi ^2$, where $\rho =2/\sqrt{3} $ is a geometrical factor
for the triangular lattice. In the thermodynamic limit, $J$ is finite
in the phase coherent state and vanishes in the incoherent state;
(ii) A domain-wall defect energy is obtained as $\Delta E_2=E_r-E_p$
, where $E_r$ is the ground state energy with reflected boundary
conditions \cite{min}, corresponding to the energy cost for a domain
wall in the vortex lattice. In presence of disorder, $\Delta E_1$ and
$\Delta E_2$ fluctuate between samples, with a distribution that can
be characterized by its moments.  Stability of the ground state against
thermal fluctuations requires that the average $[\Delta E]$, where $[ $
$]$ denotes a disorder average, is finite or increases with $L$ for the
$U(1)$ and $Z_2$ symmetries respectively. Fig. 2 shows the behavior of the
$[\Delta E_1]$ as function of $L$ for increasing dilution. For small $x$
, it increases with $L $, indicating the existence of long range phase
coherence \cite{trend}. For sufficiently large $x$ it clearly decreases
for increasing $L$, indicating a disordered phase. The change in the
behavior yields an estimate of $ x_S=0.14(1)$ . Fig. 3 shows a similar
plot for $[\Delta E_2]\,. $ The increasing trend with $L$ for small $x$
corresponds to a vortex-lattice ordered phase, which persists for a small
but finite range above $x_S$. For large $x$, it decreases with $L$, and
yields an estimate of  $x_{VL}=$ $0.17(1)\,$.  Thus $x_{VL}>$ $x_S$,
as indicated in Fig. 1. The disordered phase for $ x_{VL} < x <x_p$
can be regarded as a vortex glass, since it  lacks long range order in
the vortex lattice.

The stability of the glass phase against thermal fluctuations is
determined by the size dependence of the second moment of the energy 
excitations 
$w_i=\sqrt{[\Delta E_i^2]-[\Delta E_i]^2} \propto L^\theta $.  
Here $\theta > 0$ indicates a glass phase at nonzero temperature,
whereas $\theta < 0$ implies that arbitrarily low energy excitations
at long length scales can be thermally excited, destroying the glass
phase at any finite temperature \cite{bray}. The size dependence of $w$,
for a value of $x=0.3$ in this region \cite{glass} is shown in Fig. 4
and clearly indicates a negative $\theta $ for both $w_1$ and $w_2$,
and so the vortex glass only occurs at $T=0$ . The exponent $\nu _T$
$=1/\mid \theta \mid $ of the superconducting thermal correlation length
$\zeta $ $\propto $ $T^{-\nu _T}$ can be estimated from the slope of
$w_1$ in a loglog plot, giving $\nu _T\sim 1.9$. Interestingly enough,
this estimate is very close to the value obtained for the gauge-glass
model \cite{fisher}, suggesting a common universality class, but further
data would be necessary to check whether  $\nu_T$ is $x$- dependent.

\begin{figure}[tbp]
\centering\epsfig{file=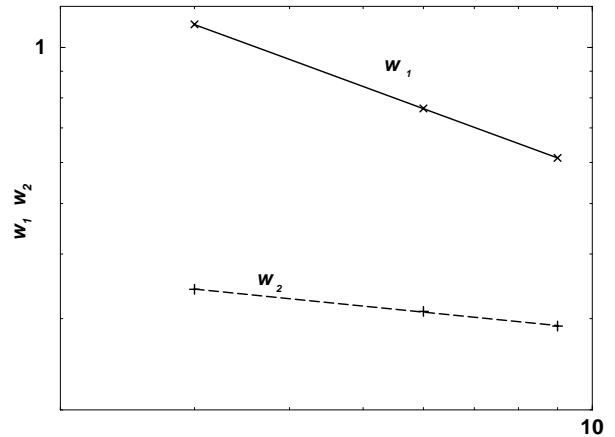,bbllx=3cm,bblly=3cm,bburx=20cm,
bbury=25cm,angle=-90,width=8cm}
\caption{Finite size behavior of the second moment of the defect energy
distribution $w_1$ and $w_2$ in the region  $x_{VL}<x<x_p$ for $x = 0.3$. The
negative slope of $\log w_1$ $\times$ $\log L$ gives an estimate of $1/\nu_T$.}
\end{figure}

At finite temperatures, thermally excited vortices and disorder
effects can significantly reduce the ordered phases for $x<x_{VL}(f)$
since bond dilution introduces correlated randomness in the flux, as in
the case of an array with disorder only in the positions of the grains
\cite{gk89}.  Unlike positional disorder, random dilution does
not explicitly affect the phase difference $\theta _i-\theta _j$ between
two superconducting grains in Eq. (1). Its relevance can be
studied through two coupled frustrated XY models
\begin{eqnarray}
H= & -\frac J2\sum_{<ij>} \big[\cos (\theta _i-\theta _j-A_{ij}) \nonumber\\
& +\cos (\phi _i-\phi _j-A_{ij}-\pi t_{ij})\big] \nonumber \\
& -h\sum_i\cos (\theta _i-\phi _i)
\end{eqnarray}
where $t_{ij}$ is $1$ or $0$ with probability $x$ and $1-x$
respectively. In the limit $h\rightarrow \infty $, the phases are coupled
$\theta _i=\phi _i$, and the original model in Eq. (1) is recovered. The
second term  has the same form as the Hamiltonian describing positional
disorder in a superconducting array in the presence of magnetic field
\cite {gk89}, with a particular bimodal distribution of $t_{ij}$. A
detailed analysis in the small $h$ limit combined with known $T \ne 0$
results \cite{physica,teitel} for $x=0$  and the above  calculations at
$T=0$, suggest the phase diagram of Fig.1.  For coupled XY models without
disorder \cite{gk89}, the coupling $h$ renormalizes to large values even
when  initially small, while the phase transitions can be described in
terms of vortices in the average phase variable $(\theta _i+\phi _i)/2$.
Guided by this, we consider initially the two XY models in Eq. (2) to
be independent, and consider the particular rational value, $f=1/2$,
where the relevant excitations, chiral domain walls and vortex charges,
are better understood \cite{physica,gk89}.  In this case, the disorder
variables act as random bonds on the chiral order parameter $\chi$,
and as random dipoles on the vortex charges.  
If the transition in the pure case is single (simultaneous disordering
of the chiral and XY-like variables), the differently-acting disorder can
thus separate the two transitions. with vortex unbinding at temperatures
below the chiral transition \cite{gk89}. In fact, Monte Carlo simulations
for the frustrated XY model on a square lattice with positional
disorder, are consistent with the splitting into two transitions
\cite{nicolides}. For the triangular lattice considered here, we  have
estimated the chiral transition temperature at  $x = x_S$, where $T_S =0$
(Fig. 1), from the peak in the chiral susceptibility and found $T_{VL}
=0.27(3)$ which can be compared with the estimated separation \cite{xu}
$\Delta T_c=0.01$ at $x=0$, if one assumes a double transition, which
clearly shows that disorder tends to separate the  transitions.
The chiral transition is expected to be in the universality class of
the random bond Ising model, where recent studies have shown that the
specific heat has a broad peak with a very weak $\log \log (T-T_c)$
divergence but the other exponents remain with the pure Ising model
values \cite{selke}. This is consistent with Monte Carlo simulations of
the frustrated XY model on a site-diluted square lattice \cite{stroud},
where it is found that the specific heat has a broad peak which does
not clearly grow with lattice size, in contrast to the undiluted case
which grows almost logarithmically.  Even when a finite coupling between
the two terms in the Hamiltonian of Eq. (2) is taken into account, the
effects of  disorder on the chiral order parameter should still remain,
since the coupling term should essentially lock equivalent vortices and
chiral variables in both phases $\theta _i$ and $\phi _i$.  For other
values of $f$, we expect similar qualitative behavior, as illustrated in
Fig. 1, but with the chiral transition replaced by the thermal disordering
transition of a vortex lattice with a higher order discrete symmetry.

Experimentally, the vortex glass phase for $ x_{VL} < x < x_p$ could be
identified  through the change in the  current-voltage characteristics
\cite{dekker} extracting the critical exponent $\nu_T$.  Another signature
would be the disappearance of ordered-phase resistance-minima at $f=p/q$
when $x$ is in the $f$-insensitive vortex-glass region $x_{VL}(f=1/2)
< x < x_p$.

\medskip
The work of EG was supported by ICTP/IAEA and FAPESP(Proc. 97/07250-8).


\begin{references}
\bibitem{trend}  This increasing trend, however, is a small size
effect. For large $L$, it should scale as $L^{d-2}$.
\end{references}
\end{document}